\documentclass[useAMS,usenatbib]{mn2e}
\usepackage{graphicx,fleqn}
\DeclareGraphicsExtensions{.eps,.ps,.eps.gz,.ps.gz,.eps.Z}
\def \msun{{\rm ~M_{\odot}}}

\def \apj{ApJ}

\def \apjs{ApJ Suppl.}
\def \aap{A\&A}

\def\ltsim{\raisebox{-.5ex}{$\;\stackrel{<}{\sim}\;$}}
\def\gtsim{\raisebox{-.5ex}{$\;\stackrel{>}{\sim}\;$}}

\newcommand{\sw}{{\it Swift}}

\def \grba{GRB~050801}

\title[
Early afterglow detection in the \sw \ observations of \grba]{ Early
afterglow detection in the \sw \ observations of \grba}
\author[]{ Massimiliano De Pasquale$^{1}$, S. R. Oates$^{1}$, M.J.
Page$^{1}$, D.N. Burrows$^{2}$
\newauthor A.J. Blustin$^{1}$, S. Zane$^{1}$, K.O. Mason$^{1}$,
P.W.A. Roming$^{2}$, D. Palmer$^{3}$,
\newauthor
N. Gehrels$^{4}$, B. Zhang$^{5}$  \\
$^{1}$ Mullard Space Science Laboratory, University College London,
Holmbury St. Mary, Dorking Surrey, RH5 6NT, UK; mdp@mssl.ucl.ac.uk \\
$^{2}$Department of Astronomy and Astrophysics, Pennsylvania State
University, 525 Davey Laboratory, University Park, PA 16802 USA \\
$^{3}$Los Alamos National Laboratories, Los Alamos, NM87545, USA\\
$^{4}$ Goddard Space Flight Center, NASA, Greenbelt, MD 20771 \\
$^{5}$ Dept. of Physics, University of Nevada, Las Vegas, NV 89154\\
}

\begin{document}

\date{Accepted...Received...}

\maketitle

\label{firstpage}

\begin{abstract}
We present results of \sw \ optical, UV and X-ray observations of
the afterglow of \grba. The source is visible over the full optical,
UV and X-ray energy range of the \sw \ UVOT and XRT instruments.
Both optical and X-ray lightcurves exhibit a broad plateau ($\Delta
t/t \sim 1$) during the first few hundred seconds after the
$\gamma$-ray event. We investigate the multiwavelength spectral and
timing properties of the afterglow, and we suggest that the
behaviour at early times is compatible with an energy injection by a
newly born magnetar with a period of a few tenths of a millisecond,
which keeps the forward shock refreshed over this short interval by
irradiation. Reverse shock emission is not observed. Its suppression
might be due to GRB ejecta being permeated by high magnetic fields,
as expected for outflows powered by a magnetar.
 Finally, the multiwavelength study allows a determination of the
burst redshift, $z=1.56$.
\end{abstract}



\section{Introduction}
\label{intro}

The \sw \ mission represents a major breakthrough for the study of
Gamma-Ray Bursts (GRBs). New bursts are detected by the Burst Alert
Telescope (BAT, \citealt{bar05}), a coded-mask imager with higher
sensitivity than that of the BATSE instrument onboard the Compton
Gamma-Ray Observatory. \sw \ can begin observing a GRB with the
X-ray telescope (XRT, \citealt{bur05}) and Ultraviolet and Optical
Telescope (UVOT, \citealt{rom05}) within $\sim1$ minute of the GRB
onset, because of \sw 's unprecedented capability of fast and
autonomous repointing. \sw \ has therefore enabled us to routinely
explore these sources from the end of the prompt emission and the
beginning of the afterglow, an epoch which was poorly known
before this mission. The observations with the XRT and UVOT
telescopes are addressing a series of key issues in GRB studies,
such as the link between the $\gamma$-ray and the afterglow
emission, the mechanisms which produce them, the duration and the
properties of the ``central engine'', and the origin of the prompt
optical emission.\\
  In this paper, we focus on \grba. Among the bursts observed by
\textit{Swift}, this GRB was the first to have an early afterglow
characterized by a peculiar phase of steady flux, both in the X-ray
and in the Optical/UV, lasting a few hundred seconds. We will show
that this distinguishing behaviour might be explained if we assume
that the burst ejecta receive an injection of energy from a
magnetar, born in the event that caused the GRB, with an initial
period of a fraction of milliseconds.\\
 \grba   could thus be another example, in addition to
GRB~051221 (\citealt{fax06}), of a GRB with a magnetar as central
engine.

\section{Prompt emission}
\label{Prompt emission}
\begin{figure*}

\includegraphics[angle=0,scale=0.5]{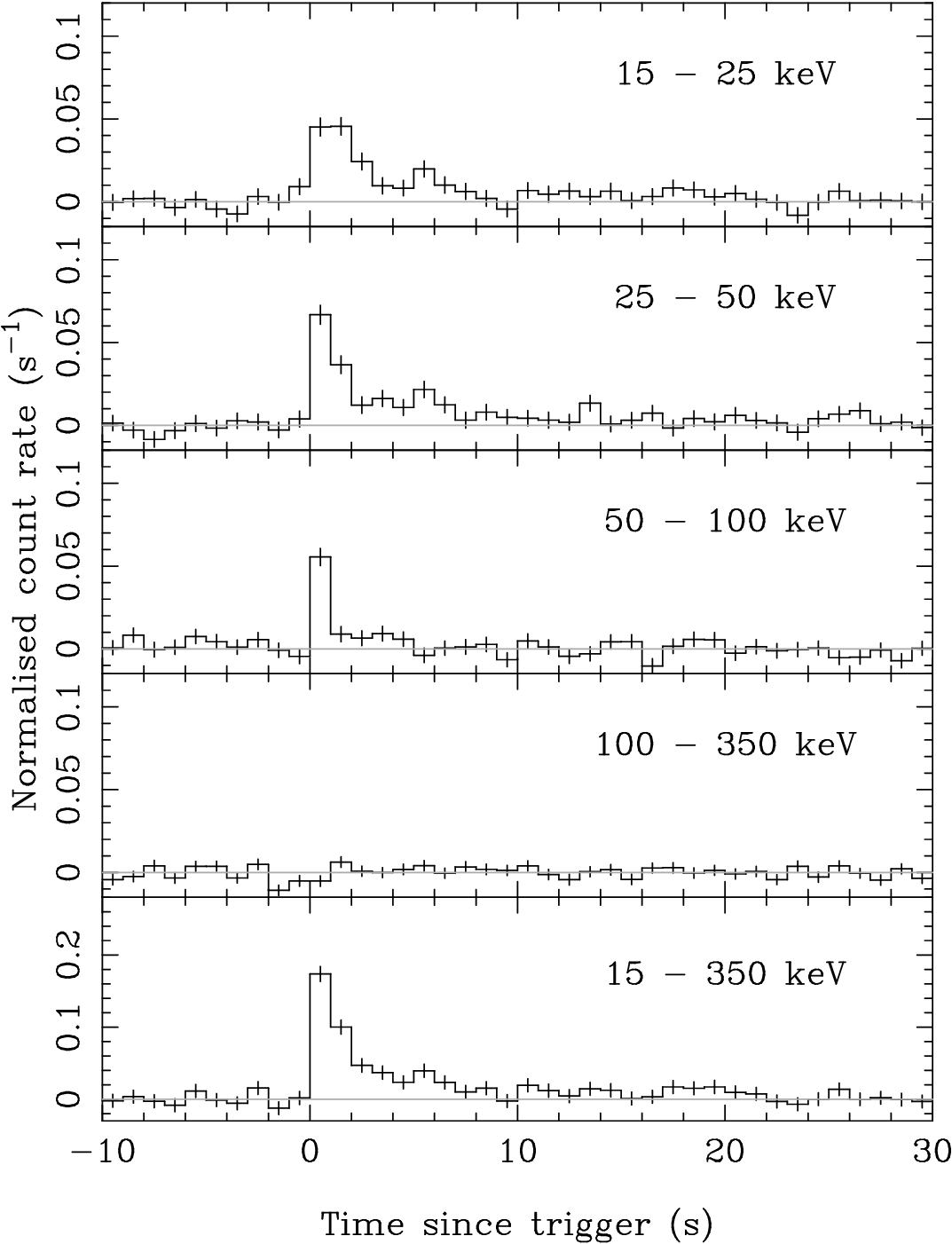}
\caption{The GRB~050801 BAT lightcurve. From top to bottom:
lightcurves in the 15-25 keV, 25-50 keV, 50-100 keV, 100-350 keV,
and 15-350 keV energy bands. Units on the Y axis are counts per
second per detector.} \label{lc_gamma}
\end{figure*}

  This burst was detected by the \sw \ satellite on August 1, 2005
at 18:28:16 UT. The $\gamma$-ray lightcurve registered by BAT
(15-350 keV) shows two peaks, with a duration t$_{90} = 20\pm2$\,s
(see Fig.~\ref{lc_gamma}). Henceforth errors are at the 68\%
confidence level, unless otherwise stated. The spectral analysis was
performed only in the 15-150 keV band, because this is the band in
which BAT has the optimal sensitivity. We tried the Band, cut-off
powerlaw and simple powerlaw models to fit the data. None of them
gave a fit significantly better than the others. However, the
parameters in the first two are poorly constrained. We have
therefore adopted the results of the simple powerlaw model fit,
which gives a spectral index $\beta=-1.00 \pm 0.16$ with a reduced
chi-squared of $\chi_{\nu} = 1.43$ with 73 d.o.f.. The total fluence
in the 15-150~keV band is $2.88\times 10 ^{-7}$ erg cm$^{-2}$, which
allows us to classify this burst as moderately faint. Interestingly,
the second peak is softer than the first one, being visible only at
energies lower than $\sim 50$keV.  In the following, we present the
X-ray, optical and UV observations performed with \sw \ and we
discuss the multiwavelength properties of the \grba's afterglow.

\section{ \emph{Swift} X-ray and optical observations of \grba}
\label{xray-opt}

 Following the BAT trigger, \grba\ was observed by the
XRT and UVOT instruments. A bright, unknown X-ray and optical source
was found inside the BAT error circle, at coordinates RA=13:36:35.4,
Dec=-21:55:42.3 (as measured with UVOT). This source subsequently
faded, and it was therefore confirmed as the afterglow of this
burst. Other prompt optical observations were successfully performed
with ROTSE (\citealt{ryk06}), and further observations were made by
\cite{fyn05a,fyn05b}, 6 hours and 30 hours after the burst onset
respectively. No radio afterglow was detected (\citealt{cameron05}).


 The UVOT began observing the field of GRB~050801 52s after the BAT
trigger. The first $\sim$9.4s exposure, taken while the spacecraft
was settling on the target, was taken through the V filter in
photon-counting (``event'') mode. Once the pointing had stabilised,
a 100s V-band finding-chart exposure in combined image and event
mode was taken. After this, the UVOT cycled through the colour
filters, firstly with 10s exposures, then 100s and finally 900s. The
later exposures were taken in Image Mode (IM) only.

 A series of images were created from the settling and finding chart
event lists. A single image was produced from the settling exposure,
and the 100s finding-chart event list was used to produce 11
time-sliced images. The first 2 of these images have 5s exposure
times, and the next 9 images are of 10s duration. Counts were
extracted from the images in all filters using a 4\arcsec\ radius
aperture at the position of the afterglow. A 25\arcsec\ radius
region, offset from the source, was used to determine the background
count rates. The source count rates were aperture-corrected to a
6\arcsec\ radius to ensure compatibility with the photometry
calibration (\citealt{poole05}), and were corrected for detector
dead-time and coincidence loss (i.e. lost counts due to multiple
photons arriving in a single region of the detector within a single
frame). The count rates were converted to magnitudes and fluxes
using, respectively, the zero-points and counts-to-flux conversions
available in the Swift CALDB. The complete log of UVOT observations
is shown in Table~\ref{tab1}.

Observations using the XRT began approximately 61s after the burst
trigger. The first exposure, taken in the Image Mode (IM), did not
show any source. Subsequently, the XRT observed in Windowed Timing
(WT) mode for approximately 20s. This observational
mode did not accumulate enough counts to produce any useful spectral
information. The lightcurve was obtained from WT event data using
\textit{Xselect} with 53 arcsecond wide extraction slices for the
source and background.

The rest of the XRT data were obtained in Photon Counting (PC) mode.
The initial PC exposure lasted approximately 880s and was found to
be piled up. Comparison of the observed point spread function of the
source to the model XRT point spread function indicated that
significant pile up is restricted to the inner 7\arcsec radius
region. Therefore source spectra and lightcurves were extracted from
an annulus with an inner radius of 7\arcsec and an outer radius of
71\arcsec. Later data did not suffer from pile up and the extraction
region was a circle of radius 70.7\arcsec. The background extraction
region was taken as a circle of radius 189\arcsec for both the piled
up and the non-piled up data. An additional fainter source was found
within the extraction region and this was excluded for both the
source and the piled up extraction regions with a radius of
2.5\arcsec radius.
 For the PC mode data we included events of grade 0-12, in which
the charge is spread over $\le 4$ pixels on the detector. The total
exposure integration time in PC mode was $2.3\times10^{4}$~s.
For the first day of observations a hot column in the XRT was close
to the centre of the source, while at later times \sw \/ was
repointed to move the source away from this artefact. For spectral
analysis we used the response matrix from the latest Swift
calibration database, CALDB 20051201, and the tasks {\sc xrtexpomap}
and {\sc xrtmkarf} were used to produce effective area files,
corrected for the hot column. The pileup correction factor was also
taken from the arf files. This correction factor was dependant upon
the distance of the burst centre to the hot pixel columns.

\begin{figure*}
\includegraphics[angle=-90,scale=0.5]{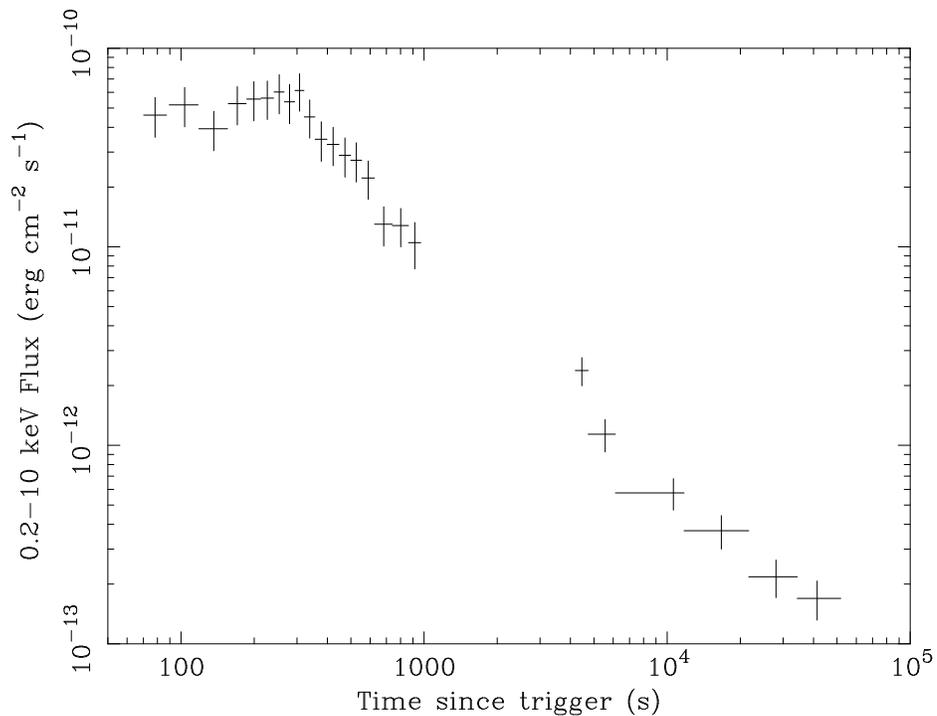}
\caption{The GRB~050801 X-ray afterglow observed with {\it Swift}
XRT in the 0.2-10~keV energy band.} \label{f1}
\end{figure*}

\begin{figure*}
\includegraphics[angle=-90,scale=0.5]{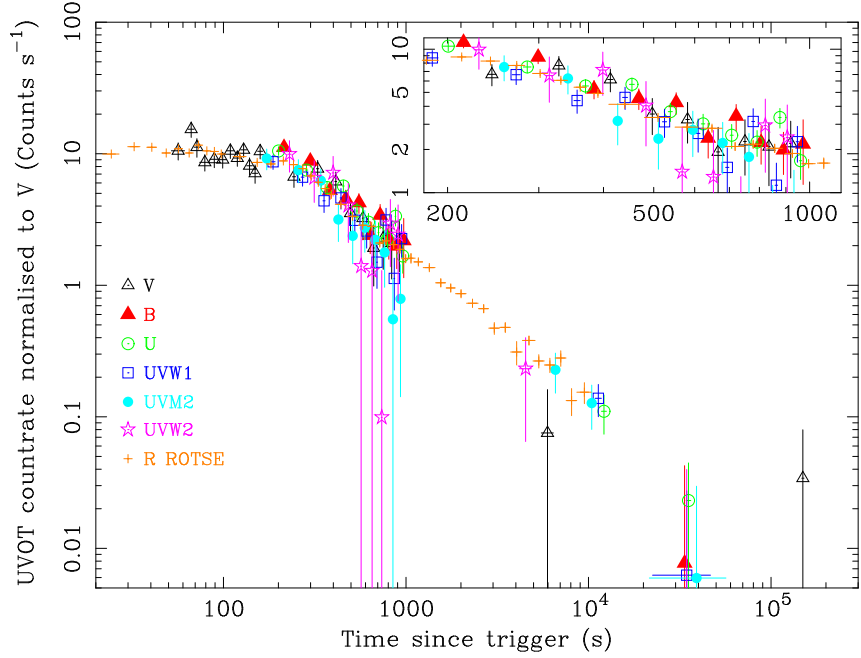}
\caption{{\it Swift} UVOT count rate of GRB~050801 in all filters,
renormalized to the V band. We also show the ROTSE R band points,
renormalized with the same criterion.} \label{f2b}
\end{figure*}

\begin{figure*}
\includegraphics[angle=-90,scale=0.5]{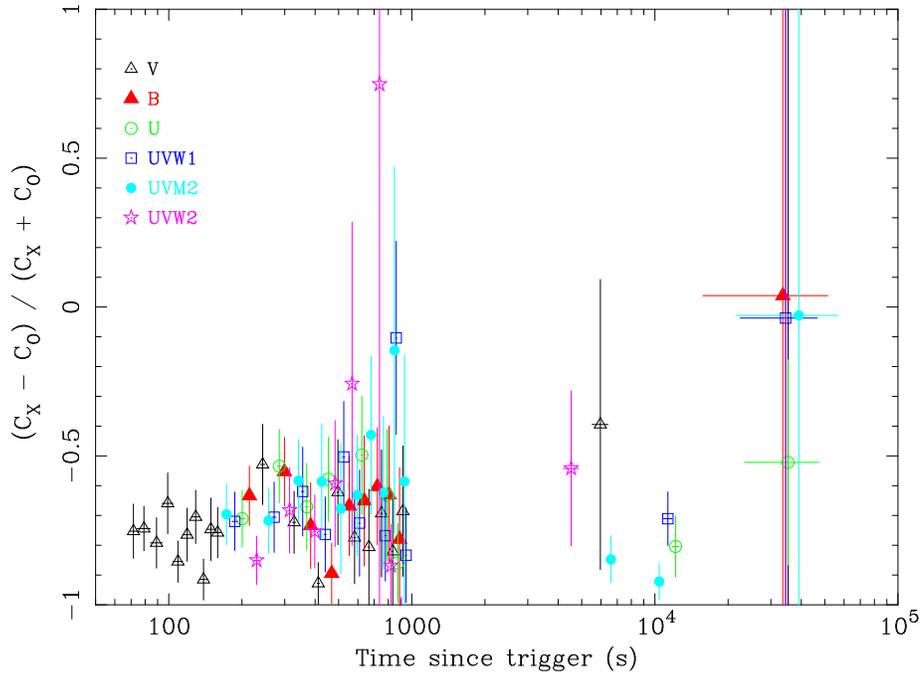}
\caption{Ratio of the {\it Swift} XRT and UVOT count rate
(renormalized). The ratio is defined as
(C$_{X}$-C$_{O}$)/(C$_{X}$+C$_{O}$), where C$_{X}$ and C$_{O}$ are
the count rates in the XRT and a given UVOT filter respectively.}
\label{f3}
\end{figure*}

Figure~\ref{f1} shows the X-ray lightcurve of \grba \ in the
0.2-10~keV energy band. The optical/UV lightcurve is shown in
Fig.~\ref{f2b}; the optical emission of \grba \ does not decline in
the first $\sim200$~s, after which it begins a systematic decay.
This behaviour is very similar for the lightcurves in all filters,
indicating no variation in the optical spectrum. We fitted the whole
dataset of optical/UV datapoints, which have been normalized to the
V band (Fig.~\ref{f2b}), with a broken powerlaw. In the following we
use the convention $F \propto t^{-\alpha}\nu^{-\beta}$, where $F$,
$t$ and $\nu$ are the flux, the time since the BAT trigger and the
photon energy respectively. The best fit values of the parameters
are: an initial decay slope $\alpha_{O,1}=0.17\pm0.09$, a break time
t$_{O,b}=230\pm15$~s and a late decay slope
$\alpha_{O,2}=1.18\pm{0.05}$.  The X-ray flux also lacks a strong
decline ($\Delta \log F < 0.5$) in the first $\sim200$~s, but decays
after this with a rate similar to that of the optical lightcurve. By
fitting this lightcurve with a broken powerlaw, we obtain an initial
decay slope $\alpha_{X,1}=-0.23 \pm 0.24$, a break time
t$_{X,b}=257^{+35} _{-24}$ s, and a late decay slope of
$\alpha_{X,2}=1.22\pm0.04$: all these values are consistent with
those obtained by fitting the optical lightcurve, within 2$\sigma$.
In fact, we find that the optical-to-X-ray flux ratio remains
remarkably constant throughout the whole observation (see
Figure~\ref{f3}). In order to formally assess that, we deconvolved
the intrinsic distribution of our data from the distribution of
measurement errors. We followed the maximum likelihood method of
Maccacaro et al. (1988) to obtain the best estimate of the intrinsic
standard deviation $\sigma_{M}$ of the optical-to-X ray flux ratio.
We got $\sigma_{OX,M} = 0.02 ^{+0.02} _{-0.01}$, which is
consistent, within 1.5 standard deviation, with 0. The $3\sigma$
upper limit on intrinsic standard deviation is 0.06.

 We also note that the level of the optical flux found by ROTSE 22s
after the trigger (\citealt{ryk06}) is consistent with the plateau
we find 70-230s after the trigger, indicating that the optical flux
was roughly constant for $\sim200$s after the end of the prompt
emission.

\begin{figure*}
\includegraphics[angle=-90,scale=0.5]{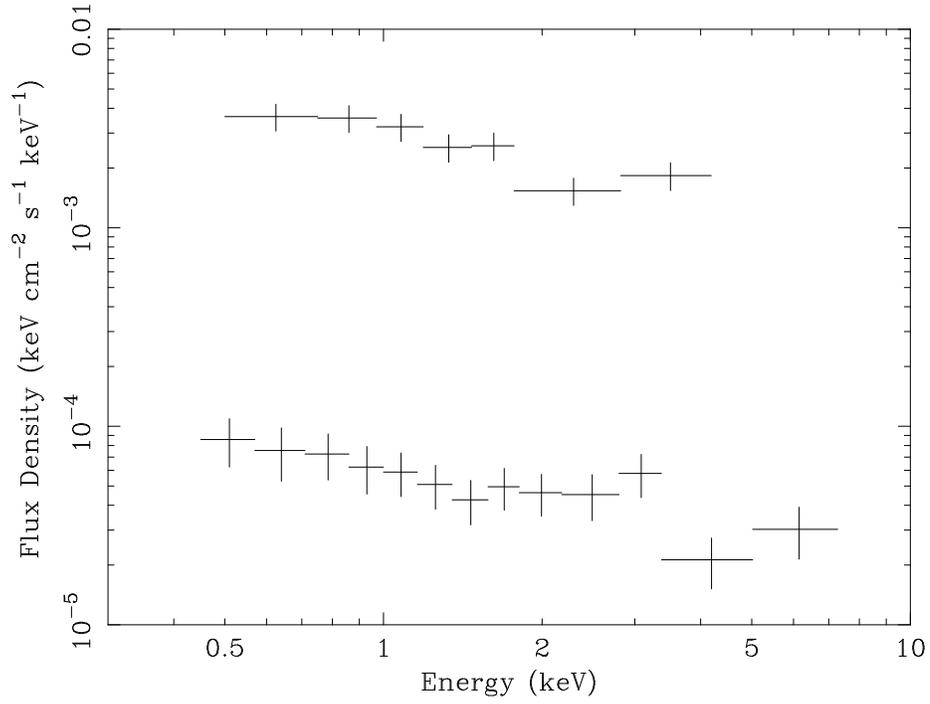}
\caption{The GRB~050801 X-ray spectrum during the piled-up phase
(time interval 90-975s, top points) and the non piled-up phase
(since 975~s, bottom points).} \label{f4}
\end{figure*}

\begin{figure*}
\begin{center}
\includegraphics[scale=0.5,angle=-90]{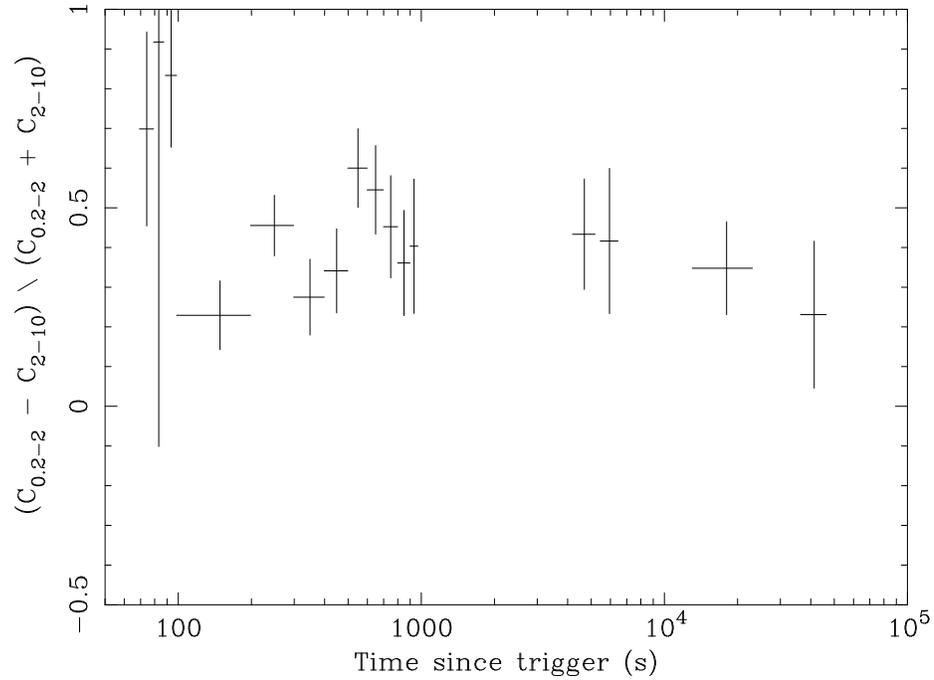}
\caption{Softness ratio of X-ray data, defined as
(C$_{0.2-2}-C_{2-10}$)/(C$_{0.2-2}+C_{2-10}$), where C$_{0.2-2}$ and
C$_{2-10}$ are the count rate in the 0.2-10 keV band and in the 2-10
keV band, respectively.} \label{f5}
\end{center}
\end{figure*}

 The 0.2-10~keV spectrum obtained with XRT is well fitted with an
absorbed powerlaw model; results are listed in Table~\ref{tab2} and
Fig.~\ref{f4}. There is no evidence for spectral evolution during
the whole of the follow-up observation. This is confirmed by the
fact that there is no change in the softness ratio, as shown in
Fig.~\ref{f5}. By means of the maximum likelihood method of
Maccacaro, we found that the best estimate of the intrinsic
dispersion of softness ratio is $\sigma_{S,M}=0.075\pm0.062$, which
is consistent with 0 at $1.1\sigma$ level.

 The similar behaviour of the X-ray and optical lightcurves, as well
as the absence of any change in the spectral properties, suggests
that these two bands both lie in the same spectral segment.
In order to test this hypothesis, and to understand the spectral
properties of the burst in more detail, we reconstructed the Spectral
Energy Distribution (SED) between these bands, spanning $\sim$4
decades in frequency.

\begin{figure*}
\includegraphics[angle=-90,scale=0.5]{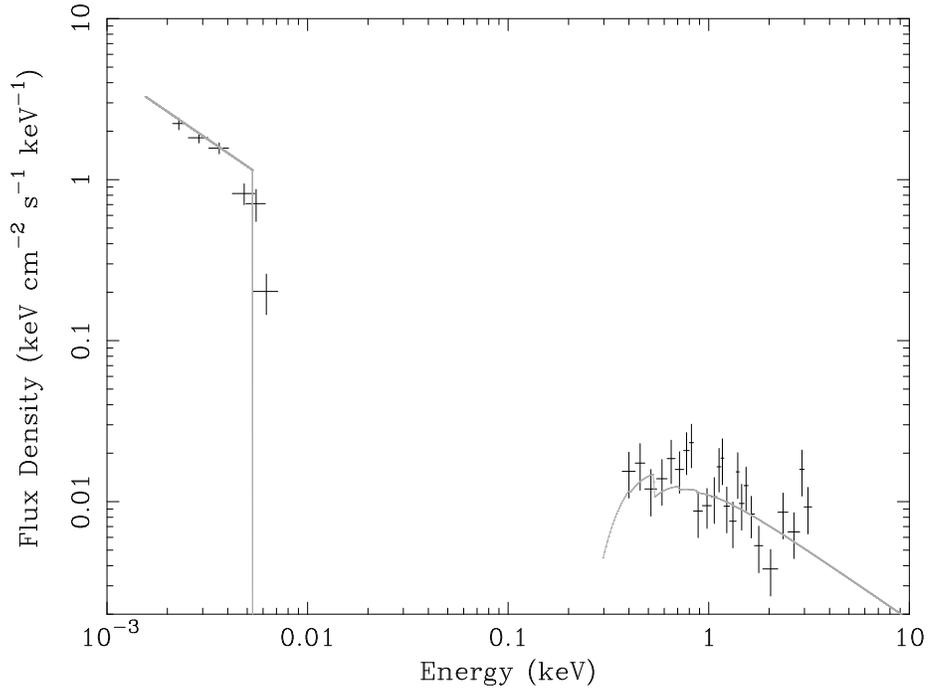}
\caption{Spectral Energy Distribution of GRB~050801, from{\it Swift}
XRT and UVOT data. The best-fit model is also plotted. In this
figure, the model and XRT datapoints were obtained by unfolding the
best fit model in \textit{XSPEC}. The optical/UV flux densities were
obtained by multiplying the UVOT countrates by the standard
countrate-to-flux conversion factors (Poole 2005). } \label{f6}
\end{figure*}

We performed a joint spectral fit of optical and X-ray data taken
between 160~s and 970~s after the burst (i.e. where both X-ray and
optical flux were highest). We fitted the spectra with a powerlaw
model, which is widely assumed to be the spectral shape of GRB
emission, and we have included the effects of photoelectric
absorption using the phabs and zphabs models in XSPEC. In our fit,
phabs is fixed at the Galactic value N$_{H}=7\times10^{20}$
cm$^{-2}$ (Dickey \& Lockman 1990), while zphabs is allowed to vary.
The optical/UV datapoints were corrected for the Galactic dust
extinction before reading into XSPEC, by using the Galactic
extinction law reported in Seaton (1979).
 Such a multicomponent model gives a rough reproduction of
the whole spectrum, with $\chi^{2}_{nu}= 38.4/27$ (see
Fig.~\ref{f6}). The best fit value of the optical to X-ray spectral
index is $\beta_{OX}=0.85\pm0.02$, which is consistent with the
spectral slope in the X-ray band, supporting a scenario in which the
X-ray and optical data lie on the same spectral segment.

Joint analysis of the optical and X-ray data has also allowed us to
derive a photometric redshift $z$ for the burst; no spectroscopic
redshift is available. Extragalactic hydrogen will cause a dropout
in the emission blueward of $91.2(1+z)$~nm, which can be detected by
\textit{Swift} UVOT if $z>1$. In the case of GRB050801, there is a
dramatic difference in flux between the UVM2 and UVW2 filters, (see
figure 7), and from the fit described above, we formally obtain
$z=1.56\pm 0.06$.  We should, however, note that the spectral model
we used includes the sharp cut-off at the Ly-limit, $91.2(1+z)$~nm,
but does not include the Ly-$\alpha$ forest, or Ly-$\alpha$
absorption from the host galaxy of the GRB. Together, these two
sources of Ly-$\alpha$ absorption will erode the spectrum between
$121.6$ and $121.6(1+z)$~nm, suggesting that the true redshift might
be slightly below the value we determined and increasing the
uncertainty on the redshift beyond the statistical errors on the
model fit. At redshift $\sim1.6$, Ly-$\alpha$ from intervening
systems is expected to absorb less than 10 per cent of the broadband
flux (Madau 2005), and will only influence the photometry in the UV
filters of the UVOT. As this attenuation is smaller than the
statistical uncertainty on the UV datapoints (15, 23 and 28 per cent
for UVW1, UVM2 and UVW2 respectively), we consider a more realistic
uncertainty for the redshift to be 0.10.


Despite these caveats, our photometric redshift should be close enough
to the actual value to permit an adequate estimation of the energy emitted by
GRB~050801 during its prompt emission phase. In calculating this energy we have
taken the K-correction into account using the method of Bloom et al. (2001),
extrapolating the flux to energies below the BAT threshold.
Assuming that the energy index does not change for $E<15$~keV, we
find the energy released over the 15-150~keV band (and in the GRB
restframe) to be $2.3 \times 10^{51}$~erg. This value puts GRB~050801
in the low energy tail of the GRB prompt energy distribution
(O'Brien et al. 2006, Nousek et al. 2006). Assuming that the spectrum of
the $\gamma$-ray emission does not change at higher
energies, we calculate the total isotropic equivalent energy of
this burst in the $1-10000$~keV band to be $9.16\times10^{51}$~erg.

\begin{table*}
\begin{center}
 \begin{tabular}{ccccc}
   \hline
Filter &   T$_{mid}$  &   T$_{range}$    & T$_{exp}$  &  Mag
\\

V      &   57     &   52$-$61         & 9.4    & 14.91 $\pm$ 0.07
\\

V      &   67     &   64$-$69         & 5.0    & 14.5 $\pm$ 0.1
\\

V      &   72     &   69$-$74         & 5.0    & 14.8 $\pm$ 0.1
\\

V      &   79     &   74$-$84         & 10.0   & 15.13 $\pm$ 0.09
\\

V      &   89     &   84$-$94         & 10.0   & 15.07 $\pm$ 0.08
\\

V      &   99     &   94$-$104        & 10.0   & 15.08 $\pm$ 0.09
\\

V      &   109    &   104$-$114       & 10.0   & 14.91 $\pm$ 0.08
\\

V      &   119    &   114$-$124       & 10.0   & 15.05 $\pm$ 0.09
\\

V      &   129    &   124$-$134       & 10.0   & 14.89 $\pm$ 0.08
\\

V      &   139    &   134$-$144       & 10.0   & 15.19 $\pm$ 0.09
\\

V      &   149    &   144$-$154       & 10.0   & 15.34 $\pm$ 0.09
\\

V      &   159    &   154$-$164       & 9.8    & 14.92 $\pm$ 0.08
\\

V      &   244    &   239$-$248       & 9.8    & 15.4 $\pm$ 0.1
\\

V      &   328    &   323$-$333       & 9.8    & 15.25 $\pm$ 0.09
\\

V      &   412    &   408$-$417       & 9.8    & 15.5 $\pm$ 0.1
\\

V      &   497    &   492$-$502       & 9.8    & 16.1 $\pm$ 0.1
\\

V      &   581    &   576$-$586       & 9.8    & 16.2 $\pm$ 0.1
\\

V      &   666    &   661$-$671       & 9.8    & 16.8 $\pm$ 0.1
\\

V      &   750    &   745$-$755       & 9.8    & 16.6 $\pm$ 0.1
\\

V      &   835    &   830$-$840       & 9.8    & 16.7 $\pm$ 0.1
\\

V      &   919    &   914$-$924       & 9.8    & 16.6 $\pm$ 0.1
\\

V      &   5965   &   5515$-$6415     & 899.8  & $>19.3$    \\

V      &   35314  &   17583$-$53044   & 2697.6 & $>20.6$    \\

V      &   149467 &   108434$-$190500 & 4211.8 & $>20.0$    \\

       &          &                   &        &            \\

B      &   215    &   210$-$220       & 9.8    & 15.32 $\pm$ 0.06
\\

B      &   299    &   294$-$304       & 9.8    & 15.57 $\pm$ 0.06
\\

B      &   383    &   379$-$388       & 9.8    & 16.13 $\pm$ 0.08
\\

B      &   468    &   463$-$473       & 9.8    & 16.29 $\pm$ 0.08
\\

B      &   553    &   548$-$557       & 9.8    & 16.37 $\pm$ 0.08
\\

B      &   637    &   632$-$642       & 9.8    & 16.99 $\pm$ 0.09
\\

B      &   721    &   717$-$726       & 9.7    & 16.60 $\pm$ 0.08
\\

B      &   806    &   801$-$811       & 9.8    & 17.06 $\pm$ 0.09
\\

B      &   890    &   885$-$895       & 9.8    & 17.19 $\pm$ 0.09
\\

B      &   972    &   970$-$973       & 3.3    & 17.1 $\pm$ 0.2
\\

B      &   33617  &   15767$-$51466   & 3599.1 & $>21.1$ \\

       &      &           &        &               \\

U      &   201    &   196$-$205       & 9.8    & 14.35 $\pm$ 0.06
\\

U      &   285    &   280$-$290       & 9.8    & 14.71 $\pm$ 0.07
\\

U      &   369    &   365$-$374       & 9.7    & 15.04 $\pm$ 0.08
\\

U      &   454    &   449$-$459       & 9.8    & 15.02 $\pm$ 0.08
\\

U      &   538    &   533$-$543       & 9.8    & 15.5 $\pm$ 0.1
\\

U      &   623    &   618$-$628       & 9.8    & 15.7 $\pm$ 0.1
\\

U      &   707    &   702$-$712       & 9.8    & 15.9 $\pm$ 0.1
\\
U      &   792    &   787$-$797       & 9.8    & 16.0 $\pm$ 0.1
\\

U      &   876    &   871$-$881       & 9.8    & 15.6 $\pm$ 0.1
\\

U      &   961    &   956$-$965       & 9.8    & 16.3 $\pm$ 0.1
\\

U      &   12159  &   11773$-$12545   & 771.3  & 19.30 $\pm$ 0.02
\\

U      &   35301  &   23342$-$47259   & 2114.0 & 20.99 $\pm$ 0.01
\\

   \hline
 \end{tabular}
\caption{UV observations of \grba \ with UVOT in different time
intervals after the BAT trigger. Errors are at 68\% C.L. For those
observations where no source was found with at least 2$\sigma$
significance, we list upper limits at $3\sigma$.} \label{tab1}
\end{center}
\end{table*}

\setcounter{table}{0}

\begin{table*}
\begin{center}
 \begin{tabular}{ccccc}
   \hline
Filter &   T$_{mid}$  &   T$_{range}$   & T$_{exp}$  &  Mag
\\

UVW1   &   187    &   182$-$192       & 9.8    & 14.7 $\pm$ 0.1
\\

UVW1   &   271    &   267$-$276       & 9.8    & 15.0 $\pm$ 0.1
\\

UVW1   &   356    &   351$-$361       & 9.8    & 15.4 $\pm$ 0.2
\\

UVW1   &   440    &   435$-$445       & 9.8    & 15.4 $\pm$ 0.2
\\

UVW1   &   525    &   520$-$530       & 9.8    & 15.8 $\pm$ 0.2
\\

UVW1   &   609    &   604$-$614       & 9.8    & 16.0 $\pm$ 0.2
\\

UVW1   &   694    &   689$-$698       & 9.8    & 16.6 $\pm$ 0.2
\\

UVW1   &   778    &   773$-$783       & 9.8    & 15.8 $\pm$ 0.2
\\

UVW1   &   862    &   858$-$867       & 9.8    & 16.9 $\pm$ 0.3
\\

UVW1   &   947    &   942$-$952       & 9.8    & 16.14 $\pm$ 0.2
\\

UVW1   &   11316  &   10866$-$11766   & 899.8  & 19.18 $\pm$ 0.04
\\

UVW1   &   34511  &   22435$-$46587  & 2699.3 & $>20.6$ \\

       &          &                   &        &               \\

UVM2   &   173    &   168$-$178       & 9.8    & 14.7 $\pm$ 0.2      \\

UVM2   &   257    &   252$-$262       & 9.8    & 14.9 $\pm$ 0.2
\\

UVM2   &   342    &   337$-$346       & 9.8    & 15.1 $\pm$ 0.2
\\

UVM2   &   426    &   421$-$431       & 9.8    & 15.9 $\pm$ 0.3
\\

UVM2   &   511    &   506$-$515       & 9.8    & 16.2 $\pm$ 0.3
\\

UVM2   &   595    &   590$-$600       & 9.8    & 16.0 $\pm$ 0.3
\\

UVM2   &   679    &   674$-$684       & 9.8    & 16.3 $\pm$ 0.3
\\

UVM2   &   764    &   759$-$769       & 9.8    & 16.5 $\pm$ 0.4
\\

UVM2   &   848    &   843$-$853       & 9.8    & 17.8 $\pm$ 0.5
\\

UVM2   &   933    &   928$-$938       & 9.8    & 17.4 $\pm$ 0.5
\\

UVM2   &   6590   &   6422$-$6759     & 337.1  & 18.7 $\pm$ 0.1
\\

UVM2   &   10415  &   9966$-$10864    & 897.3  & 19.4 $\pm$ 0.07
\\

UVM2   &   39067  &   21527$-$56607   & 2910.5 & $>20.6$ \\

       &          &                   &        &               \\

UVW2   &   230    &   225$-$235       & 9.8    & 16.1 $\pm$ 0.2
\\

UVW2   &   314    &   309$-$319       & 9.8    & 16.6 $\pm$ 0.3
\\

UVW2   &   399    &   394$-$404       & 9.8    & 16.5 $\pm$ 0.3
\\

UVW2   &   483    &   478$-$488       & 9.8    & 17.1 $\pm$ 0.3
\\

UVW2   &   568    &   563$-$573       & 9.8    & 18.2 $\pm$ 0.4
\\

UVW2   &   652    &   647$-$657       & 9.8    & 18.3 $\pm$ 0.5
\\

UVW2   &   737    &   732$-$741       & 9.8    & $>$17.9
\\

UVW2   &   821    &   816$-$826       & 9.8    & 17.4 $\pm$ 0.4
\\

UVW2   &   906    &   901$-$910       & 9.8    & 17.6 $\pm$ 0.4
\\

UVW2   &   4524   &   4188$-$4859     & 670.6  & 20.2 $\pm$ 0.05
\\

UVW2   &   34525  &   16676$-$52374   & 3599.1 & $>$21.6
\\

   \hline
 \end{tabular}
\caption{Continued.} \label{tab3}
\end{center}
\end{table*}

\begin{table*}
\begin{center}
 \begin{tabular}{cccc}
   \hline
   Segment & Energy index $\beta$ & N$_{H} \times 10^{22}$cm$^{-2}$ &
time interval (s)  \\
   \hline
  Plateau (piled up data)   &  $0.79\pm0.23$   &   -     & 90-300
\\
  Whole piled up PC data    &  $0.70\pm0.12$   & $<0.05$ & 90-975
\\
  Non-Piled up PC data      &  $0.87\pm0.23$   & $<0.06$ & 4097-52950
\\
   \hline
 \end{tabular}
\caption{Best fit parameters for the X-ray spectrum of \grba \ in
different time intervals after the BAT trigger. WT data analysis
results are not shown because of the low statistics. Errors in
$\beta$ are at 68\% C.L, while N$_{H}$ upper limits are reported at
90\% C.L. and at $z=0$. The values of N$_{H}$ represent the excess
absorption above the Galactic value of $7\times10^{20}$
(\citep{dl90})}.\label{tab2}
\end{center}
\end{table*}

\section{Discussion} \label{disc}

\subsection{Onset of the afterglow}
\label{disc_1}

\subsubsection{Constraints on the energy injection mechanism.}

The main peculiarity of the afterglow of \grba \ is the
broad plateau observed at all wavelengths from optical to X-ray
during the early phase. As we can see from Figure~\ref{f1}, the
X-ray emission is practically constant for $\sim200$~s, with a ratio
between duration $\delta t$ and time $t$ which is nearly 1. There is
no evidence for sharp decays or X-ray flares such as those observed
in several \textit{Swift} GRBs (\citealt{bur05}, \citealt{nou06}),
which are usually associated with ``internal shocks'' between the
ultra-relativistic shells emitted by the central engine.

Among the various scenarios that can explain the observed plateau
there are models in which the energy distribution over the fireball
surface is non-uniform; these include the ``patchy jet'' models
(\citealt{kp00}) and the ``two component'' jet models
(\citealt{zwh04}). In the first case, the assumption is that the
burst emission is released in a jet with large fluctuations in
angular direction, i.e. with a patchy distribution of Lorentz factor
$\gamma$. Emission from different patches becomes observable at
different times, according to their diverse values of $\gamma^{-1}$,
and this can produce features which are observed in all bands almost
simultaneously (as in \grba). It should however be noted that the
presence of large fluctuations of energy across the ejecta is still
speculative, and that in order to give rise to early ($<300$~s)
bumps, a distribution over a very narrow angular scale
($\theta\sim0.1^{\circ}$) is required. Moreover, theoretical
predictions show that in this scenario the ``bump'' should follow a
broad peak (see \citealt{zh06a}). There is no indication of this in
the available data.

 The ``two component jet'' model assumes the existence of two
components in the ejecta: one ultra-relativistic component powering
the GRB, and a second --moderately relativistic-- cocoon component. The
cocoon is decelerated later than the ultra-relativistic component
because of its lower Lorentz factor (see \citealt{zhm04},
\citealt{mr01}, \citealt{ramirez-ruiz02}). Its afterglow emission
would therefore start later and may give rise to a ``bump'' in the afterglow
lightcurve. However, in accordance with the theoretical models mentioned
above, such a feature should only be visible later than
$\sim10^{4}$~s (\citealt{zh06a}).

 The possibility we favour is that the plateau seen in both the X-ray and
optical flux corresponds to the early afterglow, i.e. to the phase
that follows the creation of an ``external shock'' after the
fireball runs into the circumburst medium. The standard afterglow
model predicts that, as a shell of ejecta interacts with the
circumburst medium, it gives rise to a ``forward shock'', which
propagates outward. In its simplest formulation, the model predicts
the production of a broad and achromatic flare, rather than a
plateau, both in the optical and X-ray bands. The non-decay of the
afterglow emission of \grba \ before 250s after the trigger might
then be explained by assuming that the central engine continues to
inject radiative energy into the fireball for a few hundreds or even
thousands of seconds after the initial explosion (late ``central
engine'' activity, \citealt{zm02}). As the emission from the central
engine stops, the afterglow decay rate steepens up to more typical
values. Assuming that the luminosity of the central engine scales as
$L \propto t^{-q}$, the predicted relation between the decay and
spectral slope is (\citealt{zm02}):

\begin{equation}\label{eq1}
    \alpha=(q/2+1)\beta +q-1 \, ,
 \end{equation}

provided that the energy band we are dealing with is between the
synchrotron peak frequency $\nu_{m}$ and the cooling frequency
$\nu_{c}$.\footnote{Note that the signs of the indices have been
reversed with respect to the original paper by \cite{zm02}, in order
to make them consistent with our convention (here both $\alpha$ and
$\beta$ are positive).} As we will show later, this is indeed the
case for both the optical and the X-ray observed frequencies. As for
the decay index, we used the weighted average of $\alpha_{X,1}$ and
$\alpha_{O,1}$, which is $\alpha_{w}=0.12\pm0.09$. For the spectral
slope, we used $\beta_{OX}=0.85\pm0.02$. From this choice we
obtained $q=0.19\pm0.07$.

This value of $q$ is consistent with 0 within 3 $\sigma$, and
interestingly such a flat luminosity distribution is expected in
models in which the gamma ray burst leads to the birth of a
millisecond pulsar with an ultra-strong magnetic field, i.e. a
so-called ``magnetar'' (see e.g. Zhang \& Meszaros 2001, Dai \& Lu
1998). If this is the case, several parameters of the newly-born
compact object can be determined. We will discuss this intriguing
scenario in more detail in \S~\ref{mag}. \\ An observationally
indistinguishable possibility is that the central engine activity is
as brief as the prompt emission itself but, at the end of the prompt
phase, the ejecta are released with different Lorentz factors (see
Rees \& Meszaros 1998, Sari \& Meszaros 2000, \citealt{pan05}). The
slowest shells would catch up with the fastest ones, once the latter
have been decelerated by interaction with the circumburst medium.
The additional energy produced when the shells shock with each other
would make the afterglow decay shallower than usual. In this case,
assuming that the mass $M$ of the ejecta follows the law $ M ( >
\gamma ) \propto \gamma^{s}$, where $\gamma$ is the shell's Lorentz
factor, one can find an effective $s$ value that mimics the effects
of continuous energy injection with the luminosity law $L\sim
t^{-q}$. Following \cite{zh06a}, this is:

 \begin{equation}\label{eq7}
    s = -(10-7q)/(2+q)
 \end{equation}

and therefore the value $q=0$ inferred above is equivalent to
$s=-4.4\pm0.4$, i.e. a very steep distribution. A steep distribution
in the shells' Lorentz factors is required in order to have
significant energy injection into the blast wave, with more energy
carried by slow shells. Furthermore, the shell injection must occur
over a suitably extended time interval in order to produce a flat
plateau instead of a flare.

As mentioned previously, both these interpretations hold if the band
where observations are taken is positioned between $\nu_{m}$ and
$\nu_{C}$. This can be inferred by means of the ``closure
relationships'', which link the value of the spectral index in a
given band with the decay index of the flux in the same band,
according to the standard afterglow scenario (for a review, see
\citealt{zhm04}). These relationships are different depending upon the
the frequency range of the observed band, the density profile of the circumburst medium
(constant as in the interstellar medium, or radially decreasing
outwards from the centre of the explosion as expected in the case of a wind
from a massive GRB progenitor), and on whether the ejecta expand spherically or in the form
of a jet.

After the end of the plateau phase, in which (according to our
interpretation) the behaviour of the afterglow is altered by late-time
energy injection, one of these relationships should be satisfied if
we assume that the afterglow behaves as theoretically predicted. In the case
of the optical and X-ray afterglow of \grba, we find that the only
closure relationship which is satisfied is $\alpha=(3/2)\beta$
--which corresponds to the case of a spherical expansion of the ejecta
into a constant density medium-- and observing frequency $\nu$
satisfying $\nu_{m} < \nu < \nu_{c}$. This also allows us to
obtain a value for $p$, the power-law index of the energy
distribution of the radiating electrons. On theoretical grounds, and
for this spectral segment, it is $\alpha=(p-1)/2$ which gives
$p=2.70 \pm 0.05$.

\subsubsection{Birth of a Magnetar?}
\label{mag}

According to Zhang \& Meszaros (2001), when the core of the stellar
progenitor collapses into a pulsar, the luminosity emitted in the form
of Poynting flux, $L_{0}$, is given by

\begin{equation}\label{eq2}
    L_{0} = I \Omega_{0}^{2} \ 2 T_{em} \sim 10^{49} B^{2}_{15} P^{-4} _{-3} R^{6}
    _{6} \, {\rm erg \, s}^{-1}\, ,
 \end{equation}
where we use the convention $Q_{x}=Q/ 10^{x}$. In the above
expression, $I$, $\Omega_{0}$, $P$, $R$ and $B$ are, respectively,
the neutron star's moment of inertia, initial angular velocity,
initial period, radius and initial magnetic field. The quantity
$T_{em}$ represents the time scale over which the emitted luminosity
is roughly constant at the level $L_{0}$; following again the same
authors, it is given by

\begin{equation}\label{eq2b}
  T_{em} = \frac{ 3c^{2} I }{ B^{2} R^{6} \Omega_0^{2}} = 2.05 \times
10^{3} (I_{45}
  B^{-2}_{15} P^{2}_{-3} R^{-6}_{6} ) \, {\rm s}\, .
\end{equation}

 After the time $T_{em}$, the model predicts a sharp cut off of
luminosity $L$, which decreases as $L\sim t^{-2}$, i.e. $q=2$. From
Zhang \& Meszaros 2002, the afterglow dynamics can be affected only
if $q<1$. Therefore, after the $T_{em}$, we have no further
modification of afterglow lightcurve due to energy injection.

 By multiplying Eq.~\ref{eq2} with \ref{eq2b}, we can estimate the
amount of energy that is produced by the newly-born neutron star and
subsequently injected into the ejecta via dipolar spin-down. We get

\begin{equation}\label{eq2c}
 L_{0} T_{em} = 2.05 \times 10^{52} P^{-2}_{-3} \, {\rm erg} \, .
\end{equation}

Our data also allow us to estimate the change in
kinetic energy, $E_{K}$, during the ``plateau'' phase. In the case where the
observing frequency is between the peak frequency and the cooling
frequency, the kinetic energy released as a function of the time is
given by (Zhang et al., 2006b)

\begin{eqnarray}\label{eq2d}
E_{K,52} & = & \left [\frac{ \nu_{18} F(\nu=10^{18} {\rm
Hz})}{6.5\times10^{-31}} \right ]^{4/(p+3)} D_{28}^{8/(p+3)}
\nonumber \\
& \times & (1+z)^{-3(p-1)/(p+3)}
t_{d}^{3(p-1)/(p+3)} f_{p}^{-4/(p+3)}  \\ & \times &
\epsilon_{B,-2}^{(-p-1)/(p+3)}
\epsilon_{E,-1}^{4(1+p)/(p+3)}
n^{-2/(p+3)} \nu_{18}^{2(p-3)/(p+3)} \nonumber
 \end{eqnarray}

where F($\nu=10^{18}$Hz) is the flux density at $\nu=10^{18}$~Hz,
$D$ is the luminosity distance, $t_{d}$ the time in the observer's
frame, $f_{p}$ is a function of the index $p$ of the powerlaw energy
distribution of radiating electrons, $\epsilon_{B}$ is the fraction
of the energy that is contained in magnetic fields, $\epsilon_{E}$
is the fraction of the energy associated with radiating electrons,
$n$ is the particle density in the surrounding medium and $\nu$ is
the frequency at which we are observing. Again, subindices indicate
normalized quantities ($Q_{x} = Q / 10^{x}$), while $t_{d}$ is in
units of days.

For a GRB at z=1.56, assuming the cosmological parameters
$H_{0}=70$, $\Omega=0.3$ and $\Lambda=0.7$, we find
$D=4\times10^{28}$~cm. For the following parameters we take fiducial values of
$\epsilon_{E}=0.3$ (\citealt{fw01}), $\epsilon_{B} =0.03$,
$n=10$ (\citealt{b03}), $\nu=10^{18}$ Hz, $p=2.7$ and $f_{p}=0.1$
(the function $f_p$ is plotted in Zhang et al.~2006b). Inserting
the time duration of the plateau [$\Delta t_d = (250-70)$~s $\approx 2.1
\times 10^{-3}$~d] into Eq.~\ref{eq2d}, and using the $0.2-10$keV flux, $F
\approx 4\times  10^{-10}$erg cm$^{-2}$s$^{-1}$, we can infer the
increase in kinetic energy during this phase. This gives $\Delta
E_{K} \approx 1.4\times10^{53}$ erg. Finally, under the simple
assumption that all the electromagnetic energy emitted by the pulsar
has been completely converted into the kinetic energy of the ejecta (we
discuss more realistic scenarios below), we have $\Delta E_{K}
= L_0 T_{em}$ and Eq.~\ref{eq2c} gives the neutron star period, $P_0
\approx 0.4 \times 10^{-3}$ s.

We can now use the value of the spin period to estimate the emitted
luminosity and the pulsar's magnetic field strength. In the cosmological rest
frame of the GRB, the plateau duration is $(250-70) \times
(1+z)^{-1} \approx 70$~s, which gives a good approximation for the
quantity $T_{em}$. From Eq.~\ref{eq2c} we therefore obtain a
luminosity $L_0 \approx 2 \times 10^{51}$ erg sec$^{-1}$ and, from
Eq.~\ref{eq2}, a stellar magnetic field of $B \approx
1.4\times10^{15}$~G (using a neutron star radius of 10~km). This
value is in the range expected for ultramagnetized neutron stars
(``magnetars''). \\

The existence of a plateau in the lightcurve also requires the
pulsar to have a significant effect on the energetics of the afterglow.
This means that the energy injected by the pulsar before the beginning
of the afterglow, which we set at a certain time $T_{0}$, must be a sizeable
fraction of the total kinetic energy that is left for the ejecta
after the prompt emission phase, $E'_{K}$:

\begin{equation}\label{eq2e}
L_{0}  T_{0} \gtsim E'_{K} \, .
\end{equation}

To test whether this condition is met, let us consider the energy
$E_{\gamma}$ radiated in $\gamma$-rays during the prompt emission.
The radiative efficiency of a GRB, defined by the relation
$\eta_{\gamma} = E_{\gamma}/(E_{\gamma} +E'_{K})$, gives a measure
of how efficiently the GRB dissipates the total energy into
radiation during the $\gamma$-ray emission phase. On theoretical
grounds (\citealt{g01}) this efficiency is not expected to be very
high, with typical values of $\eta_{\gamma} \sim 0.2$ which are in
agreement with pre-\textit{Swift} results (De Pasquale et al. 2006).
More recent analysis indicates much higher efficiencies, up to
$0.7-0.8$, although these values are hard to accommodate within the
standard ``internal shock'' emission mechanism of GRB prompt
emission. Assuming a conservative value of $\eta_{\gamma} = 0.2$,
and taking $E_{\gamma}=9.16\times10^{51}$ erg for GRB~050801 (see
\S~\ref{xray-opt}), we obtain $E'_{K} = 3.6 \times 10^{52}$ erg.
This means that the condition $L_{0} T_{0} \gtsim E'_{K}$ is met for
$T_0 \gtsim 20$s.

Unfortunately, a precise estimate of $T_{0}$, the time of the
beginning of the afterglow, is not possible. Based on our data, we
can only state that $T_{0}$ must be less than $\sim70$~s, when
\textit{Swift} began its follow-up observations. Comparing with the
BAT lightcurve, we suggest that the afterglow beginning might be
identified as the time of the second peak in the prompt emission,
which is $\sim 5$~s after the first peak. As mentioned above, this
peak is broad and clearly softer than the first peak, which is to be
expected if the first peak were due to internal shocks and the
second to afterglow emission. Such behaviour has already been
observed in the prompt emission of several GRBs, such as GRB~970228
(for a review see \citealt{fr00}). It is worth noting that, if the
afterglow begins about $\sim 5$~s after the trigger, the optical
emission detected by ROTSE can be associated with the afterglow and
connected with the flat optical lightcurve recorded by UVOT later on
(as observed). This comes with the caveat, though, that since we do
not know the behaviour of the GRB in X-rays in the interval 20-70s
after the trigger, we cannot be sure that the emission in this
period is actually due to the forward shock.

 We can, however, discuss the possibility that the energy injection,
which shapes the early afterglow curve, begins at the time of first
ROTSE observation, i.e. 20s after the trigger. In such a case, we
obtain, from Eq. 5, that the energy injected is 1.26 higher than
previously calculated. This in turn decreases the required initial
period $P$ by a factor of 1.12, i.e. we obtain $P=0.36$ms. From Eq.
3, we can infer that the magnetic field would lower by a factor
1.26, so we have $B\approx10^{15}$ G. As we can see, there is not a
big change in the values of the parameters derived.

 Finally, we shall briefly discuss the consequences of a slightly
incorrect estimate of the redshift, as previously presented. If the
true redshift is slightly above the value we used, the value of
$L_{0}$ and $T_{em}$ estimated previously should be increased and
diminished respectively. For small corrections of $z$, the product
of these two parameters (see Eq. 5) does not vary largely. In
addition, $P$ has a weak dependence on this product. Therefore, we
do not expect a large variation of $P$. For example, if the true
redshift were 10\% lower than the value we reported, the value of
$P$ should increase only by $3\%$. Conversely, if the true redshift
were $10\%$ higher than that we found, $P$ should be $3\%$ lower.
Similarly, we do not predict a noticeable variation of the magnetic
field. From Eq. 3, we can infer that the value of $B_{15}$ would
change by $\sim1\%$ for a $10\%$ variation of the redshift.
 In the same way, the condition expressed by Eq. 7 is still satisfied
in our model, for small changes in $z$ without any considerable
change of $T_{0}$.

\subsubsection{Critical issues}
\label{crit}

The discussion presented in \S~\ref{mag} is based on a number of
assumptions regarding the parameters which describe the physical
properties of the GRB. We note, however, that the dependence of the
kinetic energy of the ejecta on $\epsilon_{E}$, $\epsilon_{B}$, and
$n$ is not very strong, and so different values are not expected to
significantly affect our conclusions. On the other hand, the
assumption that all of the EM energy radiated by the pulsar is
converted into the kinetic energy of the ejecta is more
questionable. In a more realistic scenario, we would expect that
only a fraction of this EM energy is converted in this way; in order
for our scenario to be consistent, the efficiency cannot be lower
than $\sim30\%$. We certainly do not expect the conversion
efficiency to be too low, however, since the relatively low-energy
emission from the magnetar is unable to penetrate the shockwave. The
Poynting flux energy of the pulsar can therefore only propagate
outwards by conversion into the kinetic energy of the ejecta.
The constraints regarding the conversion of the EM energy may be
somewhat further eased if we assume that the ejecta outflow is moderately beamed.
While such a condition would not be in contradiction with the observations
or theory, it would further reduce the energy contribution required from the pulsar.

Other uncertainties relate to the way in which the pulsar energy
injection is estimated. We assume that the magnetic field is
constant during the phase in which energy injection is associated
with the rapid pulsar spin down, while in a more realistic scenario
the coupling between rotational and magnetic evolution should be
accounted for. Moreover, the emitted EM energy is calculated
under the simple assumption of dipolar emission in vacuum, which is
obviously a crude approximation during the first phases of neutron
star formation and evolution. Dynamo effects, the presence of
magnetospheric currents or multipolar components of the magnetic
field may complicate this simple picture.

Our overall conclusion is that, to a first order approximation, EM
energy injection from a rapidly spinning magnetar (a neutron star
rotating with a period of a few tenths of a millisecond and
possessing an ultra-strong magnetic field of order $\sim 10^{15}$~G)
can plausibly explain the observed plateau in the X-ray and optical
lightcurves of GRB~050801. Although we have not directly derived
estimates of the burst's physical parameters ($\epsilon_{E},
\epsilon_{B}$, density of surrounding medium), the assumption of
canonical values for these quantities is an adequate approximation
for the purposes of our calculations.

It is instructive to compare GRB~050801 with the short GRB~051221
\citet{fax06}. In this latter event, a flattening occurred in the
X-ray lightcurve between $3\times10^{3}$~s and $2\times10^{4}$~s
after the trigger. Optical data on this burst are rather sparse, and
so cannot help constrain the afterglow physics. In the context of
short GRBs, which are thought to arise from the coalescence of a
binary system composed of two compact objects (such as two neutron
stars or a neutron star and a black hole), it is difficult to
imagine that the fall-back accretion of part of the material onto
the central compact remnant can, hours after the coalescence, pump
the emitted energy up to $10^{51} - 10^{52}$ erg. Another scenario
was therefore put forward in which two neutron stars coalesce and
form a magnetar. In this hypothesis, the flattening of the
lightcurve was interpreted as the signature of the underlying
magnetar, whose energy injection can only significantly influence
the afterglow lightcurve after several kiloseconds post-trigger.
According to these authors, a magnetar with an initial period of
$10^{-3}$s and magnetic field of $10^{14}$~G can explain the
observed behaviour. In the case of GRB~050801, we infer values for
the period and magnetic field which are, respectively, shorter and
higher. This is because, for this burst, the magnetar energy
injection must occur over a shorter time interval, whilst the
luminosity must be higher. It should be noted, nevertheless, that
the differences in these parameters between the two GRBs are within
$\sim1$ order of magnitude.

\subsection{Absence of a reverse shock?}
\label{disc_2}

In the standard afterglow model, it is expected that the formation
of a forward shock is accompanied by that of a less energetic
``reverse shock'', which moves inward through the ejecta
(\citealt{mr97}). While the emission of the forward wave is predicted
to peak in the X-ray band (at least at the early stages of the
afterglow that we are considering here), the emission of the reverse
shock should peak in the IR-optical because of the higher density of
the ejecta that it crosses.

In the case of \grba, we do not observe any optical flares, and the
optical emission always follows the X-ray emission, with the ratio
of optical to X-ray flux remarkably constant (see Fig.~\ref{f2b}).
This suggests that the reverse shock emission is suppressed. A
possible reason may be a very high magnetic field in the ejecta, in
which case most of the energy is carried by the field itself and it
is not converted into radiation in the shells. According to
\cite{zk05}, for bursts with short duration (i.e. $t_{90} \ltsim 20
$~s), suppression of the reverse shock takes place if the
parameter $\sigma$, defined as

  \begin{equation}\label{eq3}
  \sigma = \frac{B^{e,2}}{4 \pi n m_{P} c^{2}} ,
  \end{equation}

where $B^e$ and $n$ are the magnetic field and the density of the
ejecta respectively, and $m_{p}$ is the proton mass, is larger than
100. The $\sigma$ parameter can be interpreted as the ratio between
the energy contained in the magnetic field and that in the baryonic
outflow. Since the value of $B^e$ is strongly dependent on the
assumptions made about the position of the shock radius and the
local particle density, it cannot be robustly determined. If
internal shocks occur at small radii, even a relatively low magnetic
field may prevent the reverse shock from forming. We can, however,
give an order-of-magnitude estimate of $B^e$ assuming an ejecta mass
of $10^{-4}-10^{-5}\msun$, a shock radius at the onset of the
afterglow of $r_{sh} \sim 10^{13}$cm and a spherical shell of depth
$c \delta t$ with $\delta t \sim t_{90} \sim 20$~s. This gives a
particle density of $n \sim 10^8$cm$^{-3}$ in the shell, which in
turn implies a magnetic field of $B^e\sim10^{4}$~G. Another
observational consequence of a strongly magnetized reverse shock is
the broadening of its emission profile (Zhang \& Kobayashi 2005).
This phenomenon, together with the suppression of the peak emission,
would make the reverse shock emission even less evident. It should
be noted that the hypothesis of a high magnetic field in the ejecta
may be consistent with the powering of the wind by a magnetar. Thus,
the non-detection of a reverse shock component in the emission of
this burst may be in agreement with the identification of the
central engine of GRB~050801 as a magnetar.

Other scenarios are, however, also plausible. For example, we
cannot exclude the possibility that, for unusually high density ejecta,
the reverse shock emission would have peaked in the far-IR rather than in
the IR or optical. In such a case, it would have made a negligible
contribution to the emission detected by UVOT, which is consistent with
our observations.

\section{Conclusions}

\sw \ \grba \ shows remarkable spectral and temporal features, in both
the X-ray and optical/UV bands. Instead of rapidly decaying
flux within 100-200s after the trigger, as seen in the
majority of bursts, the \grba \ emission in the X-ray and in the
optical has a plateau from $\sim 70$~s up to $\sim 250$~s after
the BAT trigger, followed by a more normal decay slope. It is
possible that the plateau actually extends back to $\sim 20$~s after
the trigger, which could explain why the first prompt optical
measurement, performed by ROTSE-III at this epoch, is consistent
with the UVOT data.

We find that the relatively flat X-ray lightcurve might be
caused by a late energy injection by the ``central engine'' to the
expanding shell, in the form of either Poynting flux or the shock dissipation
that occurs when late shells catch up with earlier ones. In the case
of energy injection due to Poynting flux,
we have shown that dipolar spinning down emission from a newly born
magnetar of initial period $P_0\simeq0.4$~s and magnetic field $B
\simeq 1.4\times10^{15}$~G can account for the X-ray flux observed
during the plateau phase and for its duration. Other models, which
involve uneven energy distribution in the ejecta or internal shock
emission are likely to be ruled out by the fact that the plateau
appears at early time and its emission is remarkably smooth.

The energy injection model requires that the afterglow has already
started by the time of the \sw \  measurements. This hypothesis is
supported by the fact that the optical-to-X-ray flux ratio during the
plateau is consistent with that observed at late times, when the
afterglow has certainly begun. We suggest that the afterglow onset
might actually be associated with the second peak detected by BAT.
This peak is very soft, as expected if it is connected with the
afterglow emission, and it occurred only $\sim 5$~s after the trigger.

Theory predicts that the forward shock in the circumburst medium,
responsible for the afterglow emission, should be accompanied by a
reverse shock that moves inwards through the ejecta. The reverse shock
emission should be brief and peaked in the optical. On the other
hand, the constancy of the optical to X-ray flux ratio observed
during this event suggests that no reverse shock emission has taken
place, at least during the \sw \ observations. The suppression of the
reverse shock emission might suggest the presence of highly
magnetized ejecta, in which most of the energy is carried by the
magnetic field rather than the shock.

Finally, a joint fit of the X-ray and optical data has allowed us to
confirm that data in both bands lay on the same spectral
segment, between the synchrotron peak and the cooling frequency, and
to determine the spectral index with high precision. Analysis of the late-time ($>300$
s) afterglow emission, based on the application of closure
relations, suggests that the GRB fireball had a spherical expansion
in a constant density environment.
More importantly, we have been able to determine the redshift of the burst as $z=1.56$.

\section{Acknowledgements}

The authors wish to thank J. Osborne, P. Boyd, P. Schady and S.
Kobayashi for the valuable comments that helped to improve the
manuscript. SZ thanks PPARC for its support through a PPARC Advanced
Fellowship.

\clearpage

\clearpage

\end{document}